\documentclass[aps,prl,twocolumn,amsfonts,showpacs,superscriptaddress]{revtex4}
\usepackage[dvips]{graphicx}
\usepackage{bm}
\usepackage{epsfig,amsopn}
\usepackage{graphicx}
\usepackage{amsmath,amssymb}
\usepackage{natbib}
\renewcommand{\section}[1]{{\par\it #1.---}}
\def\etal{et al.}

\begin{document}
\title{
Symmetry in Full Counting Statistics, Fluctuation Theorem, and Relations among Nonlinear Transport Coefficients in the Presence of a Magnetic Field
} 
\author{Keiji Saito}
\affiliation{
Graduate School of Science, University of Tokyo, 113-0033, Japan}
\affiliation{
CREST, Japan Science and Technology (JST), Saitama, 332-0012, Japan}
\author{Yasuhiro Utsumi}
\affiliation{
Condensed Matter Theory Laboratory, RIKEN, Wako, Saitama 351-0198, Japan}

\begin{abstract}
We study full counting statistics of coherent 
electron transport through multi-terminal interacting quantum-dots under 
a finite magnetic field. 
Microscopic reversibility leads to the 
symmetry of the cumulant generating function, which generalizes 
the fluctuation theorem in the context of quantum transport.  
Using this symmetry, we derive 
the Onsager-Casimir relation in the linear transport regime and 
universal relations among nonlinear transport coefficients.

\end{abstract}
\date{\today}
\maketitle

\newcommand{\mat}[1]{\mbox{\boldmath$#1$}}

{\em Full counting statistics} (FCS) has become an active topic in 
the mesoscopic physics~\cite{Levitov,QNoise,Bagrets,Braggio,Utsumi,Gogolin,BUGS,Saito,Belzig,reulet05,Gustavsson,Fujisawa}. 
FCS addresses the probability distributions of charges transmitted 
during a measurement time, $\tau$. It can characterize 
the statistical properties of quantum transport in the far-from-equilibrium regime.
Since the first paper by Levitov and Lesovik~\cite{Levitov}, 
many theories have clarified various aspects
of the distributions
~\cite{Levitov,QNoise,Bagrets,Braggio,Utsumi,Gogolin,BUGS}. 
Recently, experiments have been conducted to measure third 
current cumulants~\cite{reulet05} and the 
distributions~\cite{Gustavsson,Fujisawa}. 
However, FCS has never been applied to 
exploration of general aspects, such as
nonequilibrium thermodynamic structures in coherent electron transport.
In this paper, we discuss these general aspects by studying 
symmetries in FCS, which is valid beyond a linear response regime.

Our argument is based on the microscopic reversibility, and is related to 
the steady state {\em fluctuation theorem} (FT) in
nonequilibrium statistical mechanics \cite{evan1,GG96,wang,QF}.
FT is an important theory that holds even in the far-from-equilibrium regime. 
It is written as
$\lim_{\tau\to\infty} \ln
[P(\Delta S)/P(-\Delta S)]/\tau \! =\! I_E$, where 
$P(\Delta S)$ is the probability of entropy $\Delta S \! = \! I_E\tau$,  
produced during time $\tau$. 
This expression quantifies the probability of negative entropy,
which can be finite for a short interval of time in small systems,
as demonstrated in the context of the second law violation 
in colloidal particle experiments \cite{wang}. 
Remarkably, FT can reproduce Onsager's reciprocal relations 
and the Kubo formula ~\cite{GG96,Andrieux1} and predicts properties 
in the far-from-equilibrium regime \cite{Andrieux1}.
 
Recently FT was studied with regard to 
classical mesoscopic electron transport, 
i.e., in the sequential tunneling regime in the quantum-dots,
with Markovian approximations~\cite{Andrieux1,Esposito}. 
These works highlighted the relation between FT and FCS.
In this paper, we study the general relation between FT and FCS 
with respect to {\em coherent electron transports} in generic situations.
We consider a multi-terminal interacting 
electron cavity under a finite magnetic field. 
The validity of FT has not yet been established in this regime.
In electron transport, two thermodynamic forces 
-the thermal gradient and the bias voltage- produce entropy.
Thus, we introduce a cumulant generating function to obtain heat and 
charge transfer, and 
exactly derive its new symmetry.
The symmetry leads to a quantum version of FT and measurable universal 
relations among nonlinear transport coefficients.
They are extensions of the Onsager-Casimir relation~\cite{Onsager}.

\section{Model}
We consider a mesoscopic cavity connected to $m$ electron reservoirs. 
The total Hamiltonian consists of the reservoirs ${H}_r$ 
($r \!=\! 1, \cdots, m$), 
cavity ${H}_{d}$, interaction ${H}_{\rm int}$, and tunneling 
${H}_{T}$
\begin{eqnarray}
H
 = &
\sum_{r=1}^{m} {H}_r 
+{H}_{d}
+{H}_{\rm int}
+ {H}_{T} .
\label{hamiltonian}
\end{eqnarray}
The cavity is described as
${H}_d
=
\sum_{ij\sigma} t_{ij} \, d_{i \sigma}^\dagger d_{j \sigma}
$
where $d_{i \sigma}$ annihilates an electron with spin $\sigma$ at site $i$. 
We assume the Coulomb interaction in the cavity 
$
{H}_{\rm int}
\!=\!
\sum_{i j \sigma \sigma'}
U_{i \sigma j \sigma'} 
\,
d_{i \sigma}^\dagger 
d_{j \sigma'}^\dagger 
d_{j \sigma'}
d_{i \sigma}
/2
$. 
The Hamiltonian for the reserver $r$ is 
$
{H}_r \!=\! 
\sum_{k \sigma} \varepsilon_{rk} a_{rk\sigma}^\dagger a_{rk\sigma}$, 
where $a_{rk\sigma}$ annihilates an electron with spin $\sigma$ and 
the wave vector $k$. 
The tunneling between the reservoirs and the cavity is described as 
$H_{T} \!=\!
\sum_{rki \sigma} 
t_{rki}
\, d_{i \sigma}^{\dagger} a_{rk\sigma} + {\rm H.c.}
$.
When a magnetic field $B$ is applied, 
the elements of the hopping and tunneling matrix acquire the phases
$
t_{ij}=|t_{ij}| \exp (i \, \phi_{ij}) 
$ 
and 
$
t_{rki}=|t_{rki}| \exp (i \, \phi_{ri}) 
$. 
The phases are odd functions of the magnetic field
$\phi(-B) \!=\! - \phi(B)$. 

Our calculations follow the standard procedure with a perturbation series for 
$H_{\rm int}$. Throughout this paper, we use $\hbar=k_{\rm B}={\rm e}=1$.
The density matrix at the initial time $t \!= -\tau/2$ is assumed to be 
of the product form $
\rho_0 
\!=\! 
\prod_{s}
\rho_s ,~(s=1,\cdots, m,d)$, where $\rho_s$ is the equilibrium distribution
at temperature $T_s \!=\! 1/\beta_s$ and chemical potential $\mu_s$;
$\rho_s
\!=\! 
\exp[-\beta_s (H_s \!-\! \mu_s N_s)]/{\rm Tr}\exp[-\beta_s (H_s \!-\! \mu_s N_s)]
$. 
The operator
$N_s$ is the number operator in reservoir
$N_r \!=\! \sum_{k \sigma} a_{rk\sigma}^\dagger a_{rk\sigma}$, 
or the cavity  
$N_d \!=\! \sum_{i \sigma} d_{i \sigma}^\dagger d_{i \sigma}$.
The charge and heat current operators are defined as
${I}_{c s} \! =\! i \, [N_s,H_T]$
and 
${I}_{h s}
\! = \! 
i \, [H_s,H_T]$. 
The expression for currents ${I}_{c r}$ and ${I}_{h r}$ are
given by
\begin{eqnarray}
{I}_{c r}
=& 
-\sum_{i k \sigma} 
i \,  t_{r k i}
 \, d_{i \sigma }^{\dagger}  a_{rk \sigma}
+{\rm H.c.} \, , 
\nonumber \\
{I}_{h r}
=&
-\sum_{i k \sigma} 
i \, \varepsilon_{rk} \, 
t_{r k i} \, d_{i \sigma }^{\dagger}  a_{rk \sigma}
+{\rm H.c.} \, . \nonumber 
\end{eqnarray}

\section{Cumulant generating function}
%
We introduce the characteristic function (CF) for the transmitted charge 
and heat during time $\tau$, 
$
q_{\alpha \,   s}
\!=\!
\int_{-\tau/2}^{\tau/2}  dt \, 
{I}_{\alpha \,  s} (t)
$~$(\alpha \!=\! c,h)$: 
\begin{eqnarray}
{\cal Z}( \{ \chi_{\alpha s}  \} ; B)
=&
\left \langle
V^\dagger
e^{i {H} \, \tau} 
V^2
e^{-i {H} \, \tau} 
V^\dagger
\right \rangle, 
~~
V \!=\! \prod_s V_s, ~~~
\label{cf}
\end{eqnarray}
where 
$V_s \!=\!
\exp [-i (\chi_{hs} H_s \!+\! \chi_{cs} N_s)/2]
$. This contains the counting fields for charge and heat current 
$\chi_{cs}$ and $\chi_{hs}$, respectively.
The symbol $\langle ...\rangle$ denotes an average over the initial state.
Equation~(\ref{cf}) can be rewritten in the familiar form, 
where the counting fields play roles of fictitious 
gauge fields~\cite{Levitov,QNoise,Saito}, 
since $V_r^\dagger {H}_{T} V_r$ makes the gauge fields in
the elements of the tunneling matrix. 
Note that Eq.~(\ref{cf}) satisfies the normalization condition ${\cal Z}( \{ 0  \} ; B)\!=\! 1$. 
We use the CF and introduce the cumulant generating function (CGF)
defined at the stationary state as
\begin{eqnarray}
{\cal F}( \{ \chi_{\alpha r}  \} ; B) = &
\lim_{\tau \rightarrow \infty} 
\ln {\cal Z}( \{ \chi_{\alpha r}  \} ; B)/\tau. 
\label{eqn:cgf}
\end{eqnarray}
The CGF generates cumulants from the derivatives 
with respect to the counting fields. 
The first and second derivatives generate the average current between 
the terminal 
and the cavity, and a symmetrized current correlation expressed as
\begin{eqnarray}
\langle \! \langle
I_{c1}
\rangle \! \rangle
&\!=\!&
{ \partial
{\cal F} ( \{ 0 \}; B) \over  
\partial i \chi_{c1} } =\lim_{\tau\to\infty}
{\langle q_{c1} \rangle \over \tau } ,  \nonumber \\
\langle \! \langle
I_{c1} I_{c2}
\rangle \! \rangle
&\!=\!&
\frac{\partial^2
{\cal F} ( \{ 0 \}; B)}
{\partial i \chi_{c1} 
\partial i \chi_{c2}} 
\!=\!
\lim_{\tau \rightarrow \infty}
\!
\frac{
\langle \{ q_{c1} , q_{c2} \} 
\rangle
\!-\! 
2
\langle q_{c1}\rangle 
\langle q_{c2} \rangle
}
{2 \, \tau}.
\nonumber
\end{eqnarray}

\section{Microscopic reversibility}
So far several symmetries in CGF are known \cite{Belzig}. 
For instance, CGF is a $2 \pi$-periodic function of $\chi_{cs}$,
which is a consequence of the discreteness of the charge. 
We take the time reversal symmetry in CGF into consideration. 
Let $\Theta$ be the time reversal operator, which 
evaluates the complex conjugate for complex numbers 
and reverses the spin operator~\cite{Sakurai}. 
This time reversal operator satisfies the equations;
$
\Theta \, i \, \Theta^{-1} \!=\! -i
$
and 
$
\langle n | \, {\cal O} \, | n' \rangle
\!=\!
\langle \tilde{n}' | \, 
\Theta {\cal O}^{\dagger} \Theta^{-1} 
| \tilde{n} \rangle
$~\cite{Sakurai}. Here 
${\cal O}$ is an operator and 
$
| \tilde{n} \rangle
\!=\!
\Theta
| n \rangle$.
Calculations with these equations
and use of the definition of operator $V$
lead to the relation
\begin{equation}
{\cal Z}
( \{ \chi_{\alpha s}  \} ; B)
=
{\cal Z}( \{ -\chi_{\alpha s}+ i A_{\alpha s}  \} ; -B), 
\label{eqn:eqcf}
\end{equation}
where
$A_{cs} \!=\! \beta_s \mu_s$ 
and
$A_{hs} \!=\! -\beta_{s}$. This equality is critical in obtaining 
our central result (\ref{first}).

When stationary, no extra charge and heat accumulate 
inside the cavity.
This implies that CGF depends only on the differences between 
the reservoirs' counting fields and is independent of the cavity's 
counting field and initial state.
Below, we present its proof using 
the Schwinger-Keldysh approach~\cite{Kamenev,Utsumi,Saito}. 

The entire CGF expression consists of
a noninteracting part ${\cal F}_0$ 
and an interacting part ${\cal F}_{\rm int}$.
First we consider the noninteracting part. After a straightforward
calculation, the analytical expression of ${\cal F}_0$ is obtained as 
\begin{equation}
{\cal F}_0
( \{ \chi_{\alpha s}  \} ; B)
=
-
\frac{1}{2 \pi}
\!
\int_{-\infty}^{\infty}
\!\!
d \omega ~
\ln 
\,
{\rm det} \, \hat{g}(\omega )
+{\rm const.},
\label{cgf0}
\end{equation}
where $\hat{g}(\omega )$ is the Green function which has
site, spin, and Keldysh indices.
The normalization condition 
$
{\cal F}_0( \{ 0  \} ; B) \!=\! 0
$
must be satisfied. 
The matrix elements at $({i\sigma, j \sigma'})$ for
the inverse of the Green function 
are given in the Keldysh space as 
\begin{eqnarray}
\left[
\hat{g}^{-1}
\right]_{i\sigma, j \sigma'}
\!\! &=
\omega \, \delta_{ij} \, \hat{\tau}^3
\!
-
t_{ij} \hat{\tau}^3
\! - \!
\sum_r
\hat{\tau}^3
\hat{\Sigma}_{rij}(\omega)
\hat{\tau}^3
-
\hat{\eta}_d,
\nonumber 
\end{eqnarray}
for $\sigma \!=\! \sigma'$ and $0$ for other cases.
The self-energy and the Pauli matrix $\hat{\tau}^3$ are
\begin{eqnarray}
\hat{\Sigma}_{rij}(\omega)
=
\left(
\!
\begin{array}{cc}
{\Sigma}_{rij}^{++}(\omega)
&
{\Sigma}_{rij}^{+-}(\omega)
\\
{\Sigma}_{rij}^{-+}(\omega)
&
{\Sigma}_{rij}^{--}(\omega)
\end{array}
\!
\right), 
\;\;\;\;
\hat{\tau}^3
=
\left(
\!
\begin{array}{cc}
1& 0\\
0& -1
\end{array}
\!
\right). 
\nonumber 
\end{eqnarray}
In the wide-band limit $
\Gamma_{rij} \! = \!
2 \pi \! \sum_k \! t_{rki} \, t_{rkj}^{\ast} \delta (\omega - \varepsilon_{rk} )
$ \cite{jauho}, the causal and anti-causal components are
$
{\Sigma}_{rij}^{\pm \pm}(\omega) \!=\! 
- i \Gamma_{rij} \, [1/2 - f_{r}^+(\omega)]
$ with the Fermi/hole distribution function 
$f^\pm_r(\omega) \! = \! 1/ \{ \exp[\pm \beta_r (\omega - \mu_r)] + 1 \}$.
The lesser and greater components depend on reservoir's 
counting fields as $
{\Sigma}_{rij}^{\pm \mp}(\omega) \!=\! 
\pm i \, \Gamma_{rij} f_{r}^\pm(\omega)
\exp \{ \pm i \, (\, \chi_{hr} \, \omega \! + \! \chi_{cr} ) \}
$. The wide-band limit is not critical to our argument as long as
the reservoirs form continuum energy spectra, which cover the 
entire energy range of the cavity. 
The term $\hat{\eta}_d$ has an infinitesimal contribution,
depending on the counting fields 
and the initial state of the cavity. 
This term is crucial in the causality \cite{Kamenev},
but negligible compared to the self-energy terms of the reservoirs.
This is plausible because the steady state should not depend on the 
initial state of the cavity.
In addition, we consider the rotation of the Green function 
in the Keldysh space $\hat{R}^\dagger \hat{g}_{i \sigma j \sigma'}
(\omega ) \hat{R}$, where the operator $\hat{R}$ is defined as
\begin{eqnarray}
\hat{R}
=  \exp [ i ( \chi_{hm} \omega + \chi_{cm}) \hat{\tau}^3 /2 ] \, .
\end{eqnarray}
This transformation does not change Eq.~(\ref{cgf0}), instead it
only replaces $\chi_{\alpha r}$ 
by ${\chi}_{\alpha r}\!-\!{\chi}_{\alpha m}$ in the Green function. 
Therefore, we conclude that 
${\cal F}_0$ depends only on the differences between the 
reservoirs' counting fields. 

We now consider the interaction part ${\cal F}_{\rm int}$ 
based on the linked cluster expansions.
It is formally written as 
\begin{eqnarray}
{\cal F}_{\rm int}
\! = \!
\lim_{\tau \rightarrow \infty} 
\ln \left[ e^{i \,S_{\rm int}^+ - i \,S_{\rm int}^- }
~ e^{ i \, S_J} 
\left. \right|_{J_{i\sigma} = J^{\ast}_{i\sigma} =0} \,\right]/\tau
\end{eqnarray}
Here the functions $S_J$ and $S_{\rm int}^\pm$ are given as
\begin{eqnarray}
S_J
&=&
-
\sum_{ij \sigma}
\int_{-\tau/2}^{\tau/2} 
dt \, dt' 
\hat{J}_{i \sigma}(t)^\dagger
\hat{\tau}^3
\hat{g}_{i \sigma j \sigma}(t,t')
\hat{\tau}^3
\hat{J}_{j \sigma}(t'), \nonumber \\
S_{\rm int}^\pm
&=&
- \! 
\int_{-{\tau / 2}}^{\tau / 2}
\, dt
\, H_{\rm int} 
\!
\left[
{\delta \over \delta J_{i \sigma \pm}^* (t)} \,, 
{\delta \over \delta J_{i \sigma \pm}(t) }
\right]  , \nonumber 
\end{eqnarray}
where the Grassmann source field in the Keldysh space 
$\hat{J}_{i \sigma}
\!=\! ^t({J}_{i \sigma +},{J}_{i \sigma -})$ is used.
The function $H_{\rm int}[...]$ is obtained by substituting 
the derivatives of the Grassmann numbers for the fermion operators 
in the Hamiltonian $H_{\rm int}$.
We consider these functions in the Fourier space, and 
transform the Green function and Grassmann fields
using the rotation operator $\hat{R}$. Then 
we readily find that ${\cal F}_{\rm int}$ is 
expressed by the simple replacement of
the Green function in $S_{J}$ by the inverse Fourier transform of 
$\hat{R}^\dagger \hat{g}_{i \sigma j \sigma'} (\omega ) \hat{R}$. 
In this mathematical structure,
the charge and energy conservation in the interaction process is crucial.
To check it, let us consider the calculation for one diagram that contributes 
to the second order in ${\cal F}_{\rm int}$
\begin{eqnarray}
\frac{1}{4}
\sum
\hat{\tau}^3_{s\,s}\,
\hat{\tau}^3_{s'\,s'}
U_{i \sigma, j \sigma'}
U_{i' \sigma, j' \sigma'}
\!
\int 
\!\!
\frac{
d \omega \, d \nu \, d \nu'
}{2 \pi}
g_{i \sigma, i' \sigma}^{s s'}(\nu \!+\! \omega) \, 
\nonumber \\ \times \,
g_{i' \sigma, i \sigma}^{ s' s}(\nu) \, 
g_{j \sigma', j' \sigma'}^{s s'}(\nu') \, 
g_{j' \sigma', j \sigma'}^{s' s}(\nu' \!+\! \omega),
\label{2nd}
\end{eqnarray}
where the summations are performed over the site, spin,
and Keldysh indices $s,s'\! =\! \pm$. 
Eq.~(\ref{2nd}) does not change on the rotation 
$\hat{R}^\dagger \hat{g}_{i \sigma j \sigma'}(\omega ) \hat{R}$,
since diagrammatically the energy and charge are conserved.
Therefore we find that ${\cal F}_{\rm int}$ depends only on 
the difference of the reservoirs' counting fields. 
From Eq.(\ref{eqn:eqcf}), we obtain the symmetry in the interacting 
electron transport as
\begin{eqnarray}
{\cal F}
(
\{ 
{\bm \chi}_{\alpha r} 
\};B
) 
=
{\cal F}
( 
\{ -
{\bm \chi}_{\alpha r} 
+
i {\cal A}_{\alpha r} 
\};
-B), 
\label{first}
\end{eqnarray}
where ${\bm \chi}_{\alpha r}={\chi}_{\alpha r}-{\chi}_{\alpha m}$ 
and ${\cal A}_{\alpha r}$ represents the affinity 
${\cal A}_{\alpha r}=A_{\alpha r} \!-\! A_{\alpha m}$. 
We can check that CGFs in interacting quantum-dots in 
Refs.\cite{Bagrets,Braggio,Utsumi,Gogolin} 
satisfy Eq. (\ref{first}).

\section{Fluctuation theorem}
Eq. (\ref{first}) can be regarded as a quantum fluctuation theorem (FT). 
Let us consider a two-terminal setup and define the entropy produced as
\begin{eqnarray}
\Delta S =& {\cal A}_{c1} q_{c1} \!+\! {\cal A}_{h1} q_{h1}.
\end{eqnarray}
We fix the counting fields as  
${\bm \chi}_{c 1} {\cal A}_{c 1}^{-1}
\!=\! {\bm \chi}_{h 1} {\cal A}_{h 1}^{-1}= {\chi}$.
The entropy production 
is obtained from the derivative of CGF with respective to $\chi$.
Asymptotic form of the probability distribution 
can be obtained from the inverse Fourier transform of the CF with the variable $\chi$. 
Saddle-point analysis with the symmetry (\ref{first}) yields the relation
\begin{eqnarray}
\lim_{\tau \rightarrow \infty} 
\ln \Bigl[
{P(\Delta S;B)\over P(-\Delta S;-B)}\Bigr] /\tau 
&=& 
I_E . \label{ft}
\end{eqnarray} 
This formula generalizes FT in the quantum regime 
under a finite magnetic field. At a uniform temperature, the entropy
production is proportional to the charge current. In this case, 
Eq.(\ref{ft}) 
quantifies the probability of back flow of charge currents.
\begin{figure}
\includegraphics[width=3.0in]{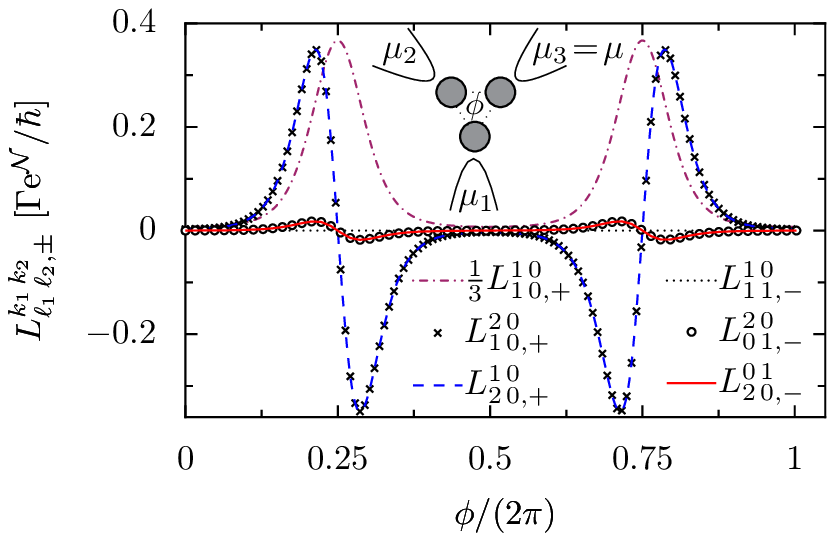}
\caption{(Color Online) The flux dependence of nonlinear transport coefficients. 
Parameters: $\varepsilon \!=\! \mu$, $t=10\Gamma $, and 
$\beta \!=\! \Gamma$.
Eqs.(\ref{3rdsym1}) and (\ref{3rdsym3}) are satisfied.
$L^{1\,0}_{1\,1,-}$ is always zero.}   
\label{fig1}
\end{figure}

\section{General relations in nonlinear transport regime}
We show that the symmetry~(\ref{first}) 
predicts general relations among nonlinear transport coefficients. 
We consider the situation that 
only the chemical 
potentials $\mu_1$ and $\mu_2$ are varied at a uniform temperature $\beta^{-1}$,
and the charge currents are measured at terminals $1$ and $2$.
We then compute cumulants of the currents,
$
\langle\langle I^{k_1}_{c1} I^{k_2}_{c2} \rangle\rangle 
=
\partial^{k_1 + k_2} 
{\cal F} ( \{ 0 \}; B) / 
\partial (\! i {\bm \chi}_{c1}\! )^{k_1} \partial (\! i{\bm \chi}_{c2}\! )^{k_2}
$. 
The nonlinear transport coefficient is defined 
in the expansion of the cumulants with respect to the affinities as
\begin{eqnarray}
L^{k_1 k_2}_{\ell_1 \ell_2}(B) 
=
\frac{
\partial^{\, \ell_1 + \ell_2 } \langle\langle I^{k_1}_{c1} I^{k_2}_{c2}\rangle\rangle
}{ 
\partial  {\cal A}_{c1}^{\ell_1} \partial  {\cal A}_{c2}^{\ell_2}
} \Bigr
|_{{\cal A}_{c1}={\cal A}_{c2}=0}.~~
\label{def_coef}
\end{eqnarray}
The affinities ${\cal A}_{cj} (j=1,2)$ are written as
${\cal A}_{cj} =\beta(\mu_j -\mu)$ with a definite chemical potential
$\mu$ for terminals $3$ to $m$. 
We symmetrize the coefficients and CGF as
\begin{eqnarray}
L^{k_1 k_2}_{\ell_1 \ell_2 ,\pm}
 &=& 
L^{k_1 k_2}_{\ell_1 \ell_2}(B) 
\,\pm \,
L^{k_1 k_2}_{\ell_1 \ell_2}(-B) 
, \nonumber \\ 
{\cal F}_{\,\pm}(\{ {\bm\chi}_{cr} \}) 
&=&
{\cal F}( \{ {\bm\chi}_{cr} \};B ) \pm {\cal F}(\{ {\bm\chi}_{cr} \}; -B) .
\nonumber 
\end{eqnarray}
From Eq.(\ref{first}), we obtain the equality
$
{\cal F}_\pm (\{ {\bm\chi}_{cr} \}) 
= 
\pm {\cal F}_{\pm} (\{ -{\bm\chi}_{cr} + i {\cal A}_{cr} \})
$. 
Note that 
CGF is a function of ${\cal A}_{cr}$ as well as ${\bm\chi}_{cr}$ \cite{Andrieux1}. 
By taking the derivatives with respect to the affinities and counting fields
for both sides of the equality of ${\cal F}_\pm $, 
we obtain the general relations among the coefficients as
\begin{eqnarray}
L^{k_1 k_2}_{\ell_1 \ell_2 ,\pm}
\! = \!
\pm \!\!\!
\sum_{n_1 =0}^{\ell_1}
\sum_{n_2 =0}^{\ell_2}
\!\!\left( \!\!
\begin{array}{c}
\ell_1 \\ n_1
\end{array}
\!\! \right)
\!
\left( \!\!
\begin{array}{c}
\ell_2 \\ n_2
\end{array}
\!\! \right) \!
(-1)^{n}
L^{k_1+n_1\, k_2 + n_2}_{\ell_1 -n_1\, \ell_2 -n_2,\pm} \, ,~
\label{relation}
\end{eqnarray}
where $n \!=\! n_1 \!+\! n_2 \!+\! k_1 \!+\! k_2$. 
With the equalities
$L^{0 \,\,0}_{\ell_1 \ell_2 ,\pm} \!=\! 0$,
the relations (\ref{relation}) can be further simplified.
We consider equations (\ref{relation}) which satisfy
${\cal N}\!=\! k_1\! +\! k_2\! + \! \ell_1\! +\!  \ell_2$.
Equations for ${\cal N} \!=\! 2$ reproduce the linear response results such as 
Kubo formula and Onsager-Casimir relations \cite{Buttiker}
\begin{eqnarray}
L^{1\,0}_{0\,1}(B)
=
L^{0\,1}_{1\,0}(-B),
\;\; 
L^{1\,0}_{1\,0}(B)
=
L^{1\,0}_{1\,0}(-B),
\end{eqnarray}
and $L^{0\,1}_{0\,1}(B) =L^{0\,1}_{0\,1}(-B)$. 
Relations beyond linear response regime can be obtained for ${\cal N} \ge 3$. 
We list some of the relations for ${\cal N}=3$ as
\begin{eqnarray}
L^{1\,0}_{2\,0,+}&=&L^{2\,0}_{1\,0,+} \, ,
\label{3rdsym1}
\\
L^{1\,1}_{0\,1,+}&=& L^{1\,0}_{0\,2,+}
=2L^{0\,1}_{1\,1,+}-L^{0\,2}_{1\,0,+} \, ,
\label{3rdsym2}
\\
L^{2\,0}_{0\,1,-}&=& L^{0\,1}_{2\,0,-}
+2L^{1\,0}_{1\,1,-} \, , 
\label{3rdsym3} \\
L^{1\,0}_{2\,0,-}&=& L^{2\,0}_{1\,0,-}/3=
L^{3\,0}_{0\,0,-}/6,
\;\;
L^{3\,0}_{0\,0,+}=0.
\label{3rdsym4} 
\end{eqnarray}
In mesoscopic experiments, large bias voltages can easily produce finite 
higher order coefficients, and the Onsager-Casimir 
relations can be violated \cite{Sanchez,Spivak,exp}.
Eqs. (\ref{3rdsym1})-(\ref{3rdsym4}) demonstrate that 
beyond the Onsager relation, universal relations exist in the nonlinear 
transport regime.
These nontrivial relations rely solely on the microscopic reversibility 
and thus are insensitive to the setup details.

\section{Three-terminal Aharonov-Bohm interferometer}
The simplest setup to demonstrate these relations would be 
a three-terminal Aharonov-Bohm (AB) ring with a threefold symmetry. 
We consider a ring consisting of three noninteracting quantum-dots, 
each of which connects to a reservoir, as shown in the inset in Fig.~\ref{fig1}. 
The Hamiltonian of the ring is given as 
$
H_d =
\sum_{\sigma}
\sum_{i=1}^{3} 
\varepsilon \, d_{i \sigma}^\dagger \, d_{i \sigma}  
-
t \, e^{i\phi/3} d_{i+1 \sigma}^{\dagger} d_{i \sigma} + {\rm H.c.}
~ (d_{4 \sigma} \!=\! d_{1 \sigma}) . 
$
The tunnel coupling is 
$\Gamma_{rij}=\Gamma \delta_{ij} \delta_{ri}$. 
The explicit form of Eq.~(\ref{cgf0}) is given by
\begin{eqnarray}
{\cal F}_0
&=&
\frac{1}{\pi} 
\! \int_{-\infty}^{\infty} \!\!\!\! d \omega
\ln \biggl 
\{ \, 1 \! + \! \sum_{j,k=1}^{3}
 f^+_j (\omega ) \, f^-_k (\omega) \,
({\rm e}^{i ( \chi_j -  \chi_k ) } \!-\! 1)
\nonumber \\ 
& & \times
[ \, 
{\cal T}_{\rm even} (\omega ) - \sum_{\ell=1}^{3}
{\cal T}_{\rm odd}(\omega)\,
\epsilon_{jk\ell} 
(1-2 f^+_{\ell} (\omega)) \, ] 
\biggl \}, 
\nonumber
\end{eqnarray}
where $\epsilon_{jk\ell}$ is the totally antisymmetric tensor. 
The transmission probability ${\cal T}_{{\rm even}/{\rm odd}}$ are
written as 
${\cal T}_{\rm even}
= 
\bar{t}^2 (1/4 + \bar{t}^2 +z^2 -2 \, \bar{t} \, z \cos \phi) /\Delta$ 
and 
${\cal T}_{\rm odd} = \bar{t}^3 \sin \phi/\Delta$, where
$z \!=\! (\omega \!-\! \varepsilon)/\Gamma$, 
$\bar{t}=t/\Gamma$ 
and 
$
\Delta
\!=\!
(z^2 \!+\! 1/4)^3 
\!+\! 
\bar{t}^2 (3/8 \!-\! 6 z^4)
\!+\! 
\bar{t}^3 z [(4 z^4 \!-\! 3) \!-\! 12 \, \bar{t}^2 ] \cos \phi
\!+\!
9 \, \bar{t}^4 (z^2 \!+\! 1/4) 
\!+\! 
4 \, \bar{t}^6 \, \cos^2 \phi
$. 
Figure~\ref{fig1} shows the linear conductance multiplied by 
$\beta^{-1}$, $L^{1\,0}_{1\,0,+}$, as well as
the nonlinear coefficients in Eqs.(\ref{3rdsym1}) and (\ref{3rdsym3}). 
It is clear that the relations (\ref{3rdsym1}) and (\ref{3rdsym3}) are satisfied. 
The overall structures depend on the temperature regions. 

\section{Two-terminal case}
Two-terminal geometry is a common setup used in experiments. 
Eqs.(\ref{3rdsym1})-(\ref{3rdsym4}) are still satisfied, but
the double script notations for
the transport coefficients are replaced by single script;
$ L^{k_1\,k_2}_{\ell_1\,\ell_2, \pm}\to (-1)^{(k_2+\ell_2 )} 
L^{k_1 + k_2}_{\ell_1 + \ell_2, \pm}$.
In general, all cumulants of
charge currents are symmetric in the magnetic field 
in noninteracting systems\cite{Konig,Sanchez,Spivak}. 
Hence, Eqs.(\ref{3rdsym3})-(\ref{3rdsym4}) cannot be measured.
However, in noncentrosymmetric interacting systems, 
the coefficient $L^{1}_{2,-}$ is generally finite \cite{Rikken}. 
In this case, Eq. (\ref{3rdsym4}) predicts a finite third current 
cumulant even in the equilibrium.
We confirmed a finite value of 
the coefficient $L^{1}_{2,-}$ and $L^{3}_{0,-} $ which is
asymmetric in the magnetic field, 
in the model of a double dot AB interferometer with noncentrosymmetric geometry 
where one of the dots is capacitively coupled to an additional gate
electrode.
We expect that the relation (\ref{3rdsym4}) 
can be observed in the setup of the experiments~\cite{exp}.

\section{Summary} 
We derived the new symmetry in full counting statistics
from the microscopic reversibility. 
This symmetry can be regarded as the quantum version of the
fluctuation theorem under a magnetic filed.
When there is no magnetic field, this reproduces the usual fluctuation
theorem without quantum corrections.
The symmetry 
leads to relations among nonlinear transport coefficients.
Eqs.(\ref{3rdsym1})-(\ref{3rdsym4})
would be measurable in present-day experiments.
We hope this work will motivate further experimental and theoretical
work on universal relations among nonlinear transport coefficients.

KS was supported by MEXT (No.~19740232). 
YU was supported by Special Postdoctoral Researchers Program of RIKEN.

\end{document}